# Momentum dependence of the energy gap in the superconducting state of optimally doped Bi$_2$(Sr,*R*)$_2$CuO$_y$ (*R*=La and Eu)

Y Okada[1*], T Takeuchi[2], M Ohkawa[3], A Shimoyamada[3], K Ishizaka[3], T Kiss[3], S Shin[3], H Ikuta[1]

[1] Department of Crystalline Materials Science, Nagoya University, Nagoya 464-8603, Japan
[2] EcoTopia Science Institute, Nagoya University, Nagoya 464-8603, Japan
[3] Institute for Solid State Physics (ISSP), University of Tokyo, Kashiwa 277-8581, Japan

E-mail: okada@mizu.xtal.nagoya-u.ac.jp

**Abstract**. The energy gap of optimally doped Bi$_2$(Sr,*R*)$_2$CuO$_y$ (*R*=La and Eu) was probed by angle resolved photoemission spectroscopy (ARPES) using a vacuum ultraviolet laser (photon energy 6.994 eV) or He I$\alpha$ resonance line (21.218 eV) as photon source. The results show that the gap around the node at sufficiently low temperatures can be well described by a monotonic *d*-wave gap function for both samples and the gap of the *R*=La sample is larger reflecting the higher $T_c$. However, an abrupt deviation from the *d*-wave gap function and an opposite *R* dependence for the gap size were observed around the antinode, which represent a clear disentanglement between the antinodal pseudogap and the nodal superconducting gap.

## 1. Introduction

One of the fundamental issues in cuprate superconductors is clarifying the nature and origin of the pseudogap [1,2]. In Bi$_2$Sr$_{2-x}$R$_x$CuO$_y$, the superconducting transition temperature $T_c$ at optimal doping decreases with the decrease in the ionic radius of *R* and the carrier range where superconductivity takes place becomes narrower [3-6]. This system, hence, provides us with different routes to unveil the relation between pseudogap and superconductivity since $T_c$ can be varied not only by changing doping but also at fixed doping [7,8].

Our previous work of angle resolved photoemission spectroscopy (ARPES) of Bi$_2$Sr$_{2-x}$R$_x$CuO$_y$ (*R*=La and Eu) using 21.218 eV photon from He I$\alpha$ resonance line has shown that the antinodal pseudogaps of Eu-doped samples (lower $T_c$) persist up to higher temperatures than La-doped samples (higher $T_c$) when compared at same doping [7]. Further, the momentum range where a clear peak was observed in the spectrum of optimally doped *R*=La and Eu samples in the superconducting state were restricted to around the node forming arc-like areas, and the length of the arc was shorter for the Eu-doped sample [8]. All these findings can be naturally understood by a competition between high temperature superconductivity and the pseudogap that is formed around the antinode [7,8].

In recent years, it has been a focus of interest if the superconducting state can be characterized by a single gap function or by two gaps that are co-existing in momentum space [1,2,9-15]. To resolve the controversy provoked by earlier studies and to increase our understanding on the nature of the gap function in the superconducting state, we studied here how the gap function of $Bi_2Sr_{2-x}R_xCuO_y$ depends on $R$ at fixed doping. If the antinodal pseudogap subsists in the superconducting state and the gap function is composed by two energy gaps, we expect that the magnitudes of these two gaps have a different $R$ dependence because of the competition between pseudogap and high-$T_c$ superconductivity, and we may disentangle the two gaps.

## 2. Experimental details

We used optimally doped single crystals of $Bi_2Sr_{2-x}R_xCuO_y$ ($R$=La and Eu) grown from polycrystalline rods that had a nominal composition of $Bi_2Sr_{1.65}R_{0.35}CuO_y$ with the same growth condition reported previously [6]. The crystals of this study were obtained by cleaving from the same domain as the samples used in the previous work using 21.218 eV photon from He Iα resonance line [8]. As shown in fig. 1(a), $T_c$ of the samples were 33 K and 18 K for $R$=La and Eu, respectively, which correspond to their optimal $T_c$ values [6,8].

The momentum dependence of the energy gap near the antinode was determined from the previously acquired data using 21.218 eV photons [8]. However, for the gap around the node, which is expected to be very small due to the low $T_c$ of the present material, the energy resolution was not sufficient. Therefore, high resolution ARPES spectra were accumulated at the Institute of Solid State Physics (ISSP) using a Scienta R4000 electron analyzer combined with an ultraviolet laser ($h\nu$=6.994 eV) as the incident light [16]. All measurements were performed at pressures below $5\times10^{-11}$ Torr and at 5 K, low enough compared to $T_c$ of the samples. The energy resolution for the laser ARPES measurement was about 2 meV.

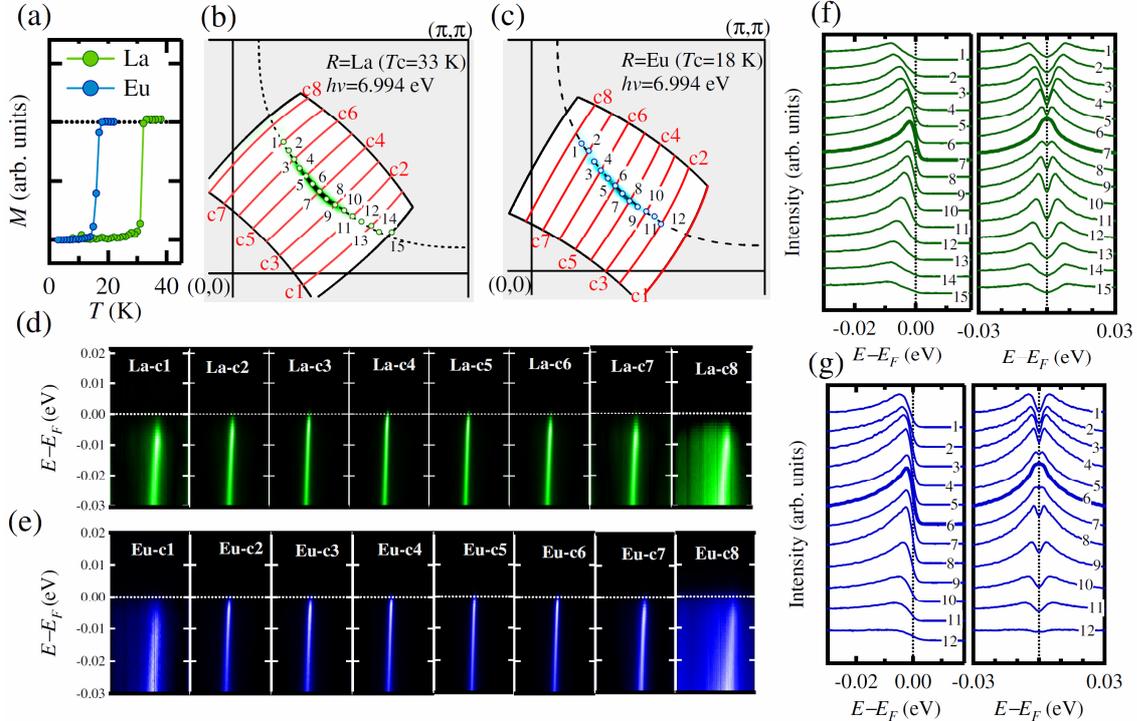

**Figure 1.** (a) Temperature dependence of magnetization (field cooling). (b), (c) Intensity maps of the ARPES spectra of the $R$=La and Eu samples, respectively. (d), (e) ARPES spectra along the cuts shown in (b) and (c). (f), (g) Energy distribution curves (EDCs) (left) and the EDCs symmetrized at the Fermi energy (right) of the $R$=La and Eu samples, respectively. The $k_F$ points where the EDCs were measured are shown in (b) and (c).

## 3. Results and discussion

Fig. 1(b) and (c) show the intensity maps of the ARPES spectra integrated within ±1 meV from the Fermi energy $E_F$ for the $R$=La and Eu samples, respectively. Fig. 1(d) and (e) show the ARPES spectra along the momentum cuts indicated in fig. 1(b) and (c). The ARPES spectra around the node of both samples are sharp while it becomes more blurred at momentum away from the node. This is a common feature of the ARPES spectra of cuprate superconductors. Fig. 1(f) and (g) show the energy distribution curves (EDCs) and those symmetrized at $E_F$ at various Fermi momentum $k_F$. The $k_F$ values where these EDCs were measured are shown in Fig. 1(b) and (c). One can see that a gap is opened except at the node and the size increases as moving toward the antinode. This momentum dependence is characteristic to a $d$-wave gap function with a point node. Interestingly, the spectral weight of the symmetrized curve tends to be more filled in for the Eu-doped sample compared with the La-doped one at the same momentum.

Fig. 2(a) shows the momentum dependence of the energy gap. The gap size here was defined as half of the peak-to-peak separation of the symmetrized EDCs. We have included in fig. 2(a) the data determined from the previously acquired ARPES spectra using 21.218 eV photon source at 8 K [8] while the symmetrized EDCs of those spectra are shown in fig. 2(b) and (c). One can see from the inset of fig. 2(a) that the gap sizes around the node of both samples change linearly as a function of $\cos 2\theta$, where $\theta$ represents the Fermi angle defined in fig. 2(b). This means that the gap around the node can be well described with a monotonic $d$-wave function at least at 5 K. The gap around the node was larger for the La-doped sample that has a higher $T_c$. All these results indicate that the nodal gap is connected to superconductivity.

On the contrary to the node, the gap size near the antinode of the Eu-doped sample was larger than that of the La-doped sample as shown in fig. 2(a), which means that the gap function has an opposite $R$ dependence around the node and near the antinode. It is hence obvious that the gap function in the superconducting state consists of two components that change in a different way when $T_c$ is altered at fixed doping. In our previous study, the antinodal gap in the normal state was found to persist up to higher temperatures for Eu-doped samples when compared with La-doped samples at same doping, which is opposite to the $R$ dependence of $T_c$ and implies that pseudogap is competing with high-$T_c$ superconductivity [7]. Here we observed that the gap size around the node changes in accordance with the change of $T_c$, while it has an opposite $R$ dependence near the antinode. Therefore, it is natural to consider that the pseudogap observed in the normal state is robust and persists also in the superconducting state.

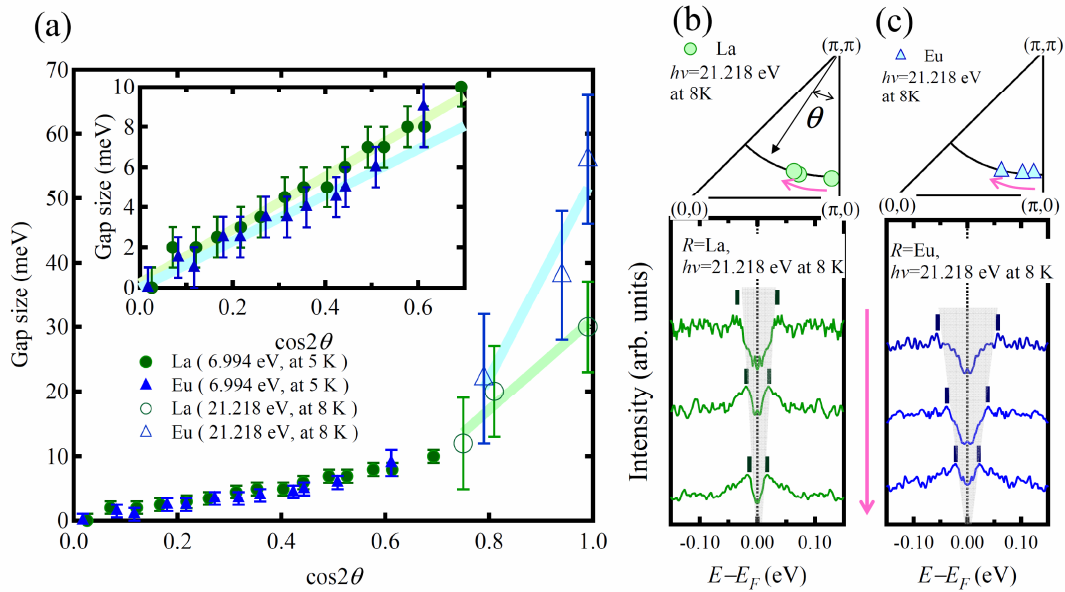

**Figure 2.** (a) Momentum dependence of the gap size in the superconducting states of the $R$=La and Eu samples. The data determined from the previously acquired spectra using 21.218 eV photon source [8] are also included. The inset shows the data in an expanded view. (b), (c) The symmetrized EDCs of the $R$=La and Eu samples measured using 21.218 eV photons. The $k_F$ positions where these spectra were taken are shown in the upper figures.

Finally we would like to mention that although the magnitude of the nodal gap changes in accordance with the change of $T_c$, the change is much smaller compared to the $T_c$ difference. This means that $T_c$ is not solely determined by the gap size around the node. Rather, the gap around the antinode shows a much larger change. A quantitative study to unveil the parameter that determines $T_c$ in this system would be necessary in the future.

### 4. Conclusions

We observed that the energy gap around the node and antinode depend on $T_c$ in an opposite way in the superconducting state of optimally doped $Bi_2Sr_{2-x}R_xCuO_y$ with $R$=La and Eu by measuring the momentum dependence of the energy gap. This result can be understood by considering two different gaps. One is related to superconductivity and is prominent in the ARPES spectra taken around the node, and the other one around the antinode is the pseudogap that competes with superconductivity.

Prior to the submission of this paper, we came aware that Hashimoto *et al.* submitted very recently a similar result about the $R$ dependence of the gap function in the superconducting state [17].